\documentstyle[preprint,aps]{revtex}

\begin{document}

\draft

\preprint {UTPT-95-18}

\title {Stellar Equilibrium and Gravitational Collapse in the
Nonsymmetric Gravitational Theory}

\author {J. W. Moffat}

\address{Department of Physics, University of Toronto,
Toronto, Ontario, Canada M5S 1A7}

\date{\today}

\maketitle

\begin{abstract}

We establish the formalism in the nonsymmetric gravitational theory
(NGT) for stellar equilibrium and gravitational collapse.
We study the collapse of a pressureless, spherically symmetric dust cloud.
By assuming that the
interior solution is smoothly matched at the surface of the star to the
quasi-static, spherically symmetric vacuum solution,
we find that the star does not collapse to a black hole. It is anticipated that
the finalcollapsed object will reach a state of equilibrium, and will emit
thermal,
gravitational and other
forms of radiation, although the radiation may be emitted only in small
amounts if the red shift from the surface of the compact object is large. No
Hawking
radiation is emitted and the information loss problem can be resolved at the
classical
level.

\end{abstract}
\vskip 0.3 true in
{}From its {\it seeming} to me -- or to everyone -- to do so, it doesn't
follow that it is so. What we can ask is whether it can make sense to
doubt it.
\vskip 0.2 true in
L. Wittgenstein, {\it On Certainty}

\narrowtext

\section{Introduction}

Recently, a new perturbatively consistent version of the nonsymmetric
gravitational theory (NGT) has been
formulated\cite{Moffat1,Moffat2,Moffat3,LegareMoff1}.
The linear approximation yields a theory free of ghost poles and tachyons
and a Hamiltonian that is bounded from below\cite{Moffat1,Moffat2}. The flux
of gravitational energy at asymptotic infinity is positive definite for a
finite value of
the range parameter $a=\mu^{-1}$. An analysis of spherically symmetric systems
in the new NGT has been carried out by Clayton\cite{Clayton}.

It was conjectured that on the basis of a static spherically symmetric
vacuum solution of the NGT field equations, in the long-range approximation,
$a \gg 2M$, no black holes would form during the collapse of a star with
a mass greater than the Chandrasekhar mass limit\cite{CornMoff1,CornMoff2}.
In order to establish the correctness of this conjecture, it is necessary to
study the situation of the physical collapse of a star containing matter as
the star's pressure falls to zero.

In order to make the solving of the field equations manageable, we must
make some simplifying approximations. We assume that the direct coupling to
the skew part of the nonsymmetric source tensor is small and can be
neglected during the collapse. We also make some simplifying approximations
about the nature of the skew symmetric contributions to the field equations.
With these approximations, we find that if the interior solution of the star
and the exterior
quasi-static solution are matched, the star collapses without forming an
event horizon (black hole) and that the collapse can stop at some time,
$t=t_0$, before it forms a state of singular energy density.

The static vacuum (Wyman) solution contains no event horizons and the
non-Riemannian
geometry is singularity-free in physical spacetime. The NGT does not
possess a Birkhoff theorem\cite{Moffat2,Clayton}, but it is reasonable
to suppose that the final state of collapse will, under physically realistic
conditions,
tend asymptotically with time towards a pseudo-stationary final state.
This state is
one in
which spacetime is invariant under an isometry group generated by a
Killing vector
field which is timelike for $0 < r < \infty$. One anticipates that
nonstationary motions will in general be damped out by gravitational
radiation,
viscosity, etc.  If such a stationary, massive final collapsed object forms,
it will for
practical purposes be indistinguishable from the observed massive ``black
holes"
reported abundantly in the literature.  The final collapsed object in NGT
will radiate
thermal radiation, gravitational radiation, etc., but not Hawking radiation,
since the
latter is a unique feature of a black hole event horizon. This could
eliminate the problem of
information loss associated with black holes in
GR\cite{Moffat3,Hawking}.

We begin, in Section II, with a presentation of the NGT field equations
and
the conservation laws. Then, in Section III, we provide a derivation of the
motion of
test particles, based on the fluid energy momentum tensor. In Section IV,
we study the formalism for a spherically symmetric system and derive the
time
dependent field
equations for such a system. In Sections V and VI, we develop the
formalism for
stellar equilibrium and gravitational collapse and in Section VII, we carry
out an
analysis of the field equations for the gravitational collapse of a
spherically
symmetric, pressureless dust cloud. This model of collapse is the
equivalent, in
NGT, of the Oppenheimer-Snyder collapse model in
GR~\cite{OppySnyder}.
In Section VIII, we study the static spherically symmetric vacuum
solution, and
discuss
the dual roles of the Riemannian and non-Riemannian geometries in
NGT.
In Section IX, we study the matching of the interior and the exterior
metrics during collapse.
Finally,
in Section X, we summarize the results of this paper and discuss
the claim made by Burko and Ori\cite{Burko,CornMoff3} that black
holes are anticipated in NGT collapse, using the linear approximation of
an expansion
of
the NGT field equations about the Schwarzschild solution of GR.

\section{Structure of the Nonsymmetric Gravitational Theory}

The NGT is a geometric theory of gravity based on a nonsymmetric field
structure with a nonsymmetric fundamental tensor $g_{\mu\nu}$,
defined by
\begin{equation}
g_{\mu\nu} = g_{(\mu\nu)} + g_{[\mu\nu]}.
\end{equation}
The affine connection coefficients, $ \Gamma^\lambda_{\mu\nu} $,
are also nonsymmetric:
\begin{equation}
\Gamma^\lambda_{\mu\nu}=\Gamma^\lambda_{(\mu\nu)}
+\Gamma^\lambda_{[\mu\nu]}.
\end{equation}
We define the inverse tensor $ g^{\mu\nu} $ by the relation
\begin{equation}
\label{contraction}
g^{\mu\nu} g_{\mu\alpha} = g^{\nu\mu} g_{\alpha\mu} =
\delta^\nu_\alpha .
\end{equation}

The NGT Ricci curvature tensor $ R_{\mu\nu}(W) $ is given by
\begin{equation}
\label{eq:Ricci_in_W}
R_{\mu\nu}(W) = W^\beta_{\mu\nu,\beta}
- \frac{1}{2} ( W^\beta_{\mu\beta,\nu} + W^\beta_{\nu\beta,\mu} )
- W^\beta_{\alpha\nu} W^\alpha_{\mu\beta}
+ W^\beta_{\alpha\beta} W^\alpha_{\mu\nu} ,
\end{equation}
where the $ W^\lambda_{\mu\nu} $ are the
unconstrained nonsymmetric connection coefficients, defined in terms
of the affine connection coefficients through the relation:
\begin{equation}
\label{eq:definition_of_W}
W^\lambda_{\mu\nu} = \Gamma^\lambda_{\mu\nu}
- \frac{2}{3} \delta^\lambda_\mu W_\nu ,
\end{equation}
where $ W_\mu = W^\alpha_{[\mu\alpha]} $.
It follows from (\ref{eq:definition_of_W}) that
$ \Gamma_\mu = \Gamma^\lambda_{[\mu\lambda]} = 0 $.
The NGT Ricci scalar is given by $ R(W) = g^{\mu\nu}R_{\mu\nu}(W)
$.

The NGT field equations take the form:
\begin{eqnarray}
\label{NGTfieldequations}
G_{\mu\nu}(W) + \Lambda g_{\mu\nu}+S_{\mu\nu} & =& 8 \pi
T_{\mu\nu} ,\\
\label{gskewequation}
{{\bf g}^{[\mu\nu]}}_{,\nu}&=&-\frac{1}{2}{\bf
g}^{(\mu\sigma)}W_\sigma,
\end{eqnarray}
where
\[
G_{\mu\nu}=R_{\mu\nu}-\frac{1}{2}g_{\mu\nu}R,
\]
and
\[
S_{\mu\nu}= \frac{1}{4} \mu^2
C_{\mu\nu}-\frac{1}{6}P^*_{\mu\nu}.
\]
Moreover,
\[
C_{\mu\nu}=\frac{1}{2} g_{\mu\nu} g^{[\alpha\beta]}
g_{[\beta\alpha]}
 +  g^{[\alpha\beta]} g_{\mu\alpha} g_{\beta\nu}+ g_{[\mu\nu]},
\]
and $P^*_{\mu\nu}=P_{\mu\nu}-{1\over 2}g_{\mu\nu}P$ with
$P_{\mu\nu}
=W_\mu W_\nu$. $\Lambda$ and $\mu$ denote the cosmological
constant and a
``mass"   parameter associated with the antisymmetric field
$g_{[\mu\nu]}$,
respectively.

We can write the field equations (\ref{NGTfieldequations}) in the form:
\begin{equation}
\label{NGT_alteqs}
R_{\mu\nu}(W)=\Lambda g_{\mu\nu}+8\pi( T_{\mu\nu}-{1\over
2}g_{\mu\nu}T
-{1\over 32\pi}\mu^2I_{\mu\nu}+{1\over 48\pi} P_{\mu\nu}),
\end{equation}
and $T=g^{\mu\nu}T_{\mu\nu}$. Also, we have
\begin{equation}
I_{\mu\nu}=C_{\mu\nu}-{1\over 2}g_{\mu\nu}C
=g^{[\alpha\beta]}g_{\mu\alpha}g_{\beta\nu}+{1\over 2}g_{\mu\nu}
g_{[\alpha\beta]}g^{[\alpha\beta]}+g_{[\mu\nu]},
\end{equation}
where $C=g^{\mu\nu}C_{\mu\nu}$.

In empty space, the field equations (\ref{NGT_alteqs}) become:
\begin{equation}
R_{\mu\nu}(\Gamma)={2\over 3}W_{[\nu,\mu]}+\Lambda g_{\mu\nu}
-{1\over 4}\mu^2I_{\mu\nu}+{1\over 6} P_{\mu\nu},
\end{equation}
where
\[
R_{\mu\nu}(\Gamma)=\Gamma^\beta_{\mu\nu,\beta}
-\frac{1}{2}(\Gamma^\beta_{(\mu\beta),\nu}+\Gamma^\beta_{(\nu\beta),\mu})
-\Gamma^\beta_{\alpha\nu}\Gamma^\alpha_{\mu\beta}+\Gamma^\alpha_{\mu\nu}
\Gamma^\beta_{(\alpha\beta)}.
\]

{}From the variational principle and the general covariance of the
Lagrangian density,
we can obtain the four Bianchi identities:
\begin{equation}
\label{Bianchi}
\frac{1}{2}(G_{\rho\nu}(\Gamma){\bf g}^{\sigma\nu}
+G_{\nu\rho}(\Gamma){\bf g}^{\nu\sigma})_{,\sigma}
+\frac{1}{2}{\bf G}_{\tau\nu}(\Gamma){g^{\tau\nu}}_{,\rho}=0.
\end{equation}
The Bianchi identities lead to the four conservation
laws\cite{LegareMoff1}:
\begin{equation}
\label{eq:energy-momentum_conserve}
g_{\mu\lambda}{{\bf T}^{\mu\rho}}_{,\rho}
+ g_{\lambda\mu}{{\bf T}^{\rho\mu}}_{,\rho}
+2[\mu\nu, \lambda]{\bf T}^{\mu\nu} = 0 ,
\end{equation}
where
\[
[\mu\nu, \lambda]=\frac{1}{2}(g_{\mu\lambda,\nu} +
g_{\lambda\nu,\mu}
-g_{\mu\nu,\lambda}).
\]
This is known as the generalized law of energy-momentum conservation
in NGT.

\section{Equations of Motion of Test Particles}

Let us consider the equations of motion of  test particles derived from the
conservation
law (\ref{eq:energy-momentum_conserve}). We shall assume that the
particle is
confined to a tube
$\Sigma$, whose linear cross section dimensions are small compared to
the length characterizing the gradient of the background metric. The
fundamental
tensor $g_{\mu\nu}$ consists of a piece $g_{\mu\nu}^{(0)}$
corresponding to the
continuous field at points along the world line of the test particle, and the
part
$\delta g_{\mu\nu}$ that describes the correction to the background
$g_{\mu\nu}^{(0)}$ field
due
to the test particle\cite{Papapetrou}. We can then write
\begin{equation}
g_{\mu\nu}=g_{\mu\nu}^{(0)}+\delta g_{\mu\nu}.
\end{equation}
To the first order of approximation, we keep in the field equations only
the terms
which are
linear in $\delta g_{\mu\nu}$. The energy-momentum tensor associated
with the test
particle is $\delta T^{\mu\nu}$. It is assumed that the energy-momentum
tensor
$T^{(0)\mu\nu}$, associated with the background metric
$g^{(0)}_{\mu\nu}$,
vanishes
inside as well as near the test particle. We adopt the notation:
\begin{equation}
{\cal T}^{\mu\nu}=\sqrt{-g}\delta T^{\mu\nu}.
\end{equation}

The energy-momentum tensor for a fluid is
\begin{equation}
\label{eq:energytensor}
T^{\mu\nu}=(\rho+p)u^\mu u^\nu - pg^{\mu\nu}+K^{[\mu\nu]},
\end{equation}
where $u^\mu=dx^\mu/ds$ is the four-velocity of a fluid element,
normalized
so that
\begin{equation}
\label{velocitynorm}
g_{\mu\nu}u^\mu u^\nu=g_{(\mu\nu)}u^\mu u^\nu=1.
\end{equation}
In the test particle limit, $p\rightarrow 0$ and we get
\[
\delta T^{[\mu\nu]}=K^{[\mu\nu]}.
\]

The equations of motion of the test particle are then given by
\begin{equation}
{{\cal T}^{(\mu\nu)}}_{,\nu}+\left\{{\mu\atop
\alpha\beta}\right\}^{(0)}
{\cal T}^{(\alpha\beta)}={\bf f}^\mu,
\end{equation}
where
\begin{equation}
f^\mu=s^{(0)\mu\alpha}(g^{(0)}_{[\beta\alpha]}{K^{(0)[\beta\nu]}}_{,\nu}
+K^{(0)[\rho\nu]}[[\rho\nu],\alpha]^{(0)}),
\end{equation}
and we have used the inverse symmetric tensor $s^{\mu\nu}$ defined by
\[
s^{\mu\nu}g_{(\sigma\nu)}=\delta^\mu_\sigma
\]
to raise suffixes. Moreover,
\begin{equation}
\label{eq:Christoffel}
\left\{{\lambda\atop \mu\nu}\right\}={1\over 2}s^{\lambda\rho}
\left(s_{\mu\rho,\nu}+s_{\rho\nu,\mu}-s_{\mu\nu,\rho}\right).
\end{equation}

Let us define the proper mass of the test particle by the equation
\[
\int d^3x {\cal T}^{00}=u^0\int
d^3x\sqrt{-g}\rho_0=m_t\frac{dt}{d\tau},
\]
where $\rho_0$ is the proper mass density , $m_t$ denotes the
test particle mass
and $d\tau$ is the proper time along the world line. The equation of
motion
now takes the form:
\begin{equation}
\label{motion}
\frac{d}{d\tau}(m_t u^\mu)+m_t\left\{{\mu\atop
\alpha\beta}\right\}^{(0)}
u^\alpha u^\beta=m_t{\tilde f}^\mu,
\end{equation}
where we have defined
\[
f^\mu=m_t {\tilde f}^\mu.
\]
We have
\[
\frac{d}{d\tau}(m_t u^\mu)=m_t\frac{d u^\mu}{d\tau}+\frac{d
m_t}{d\tau}u^\mu.
\]
{}From the condition:
\[
g_{(\mu\nu)}^{(0)}u^\mu u^\nu=1,
\]
we get
\[
\frac{1}{2}(g^{(0)}_{(\mu\nu)}u^\mu u^\nu)_{\vert\sigma}
=g^{(0)}_{(\mu\nu)}u^\nu{u^\mu}_{\vert\sigma}=u_\mu
{u^\mu}_{\vert\sigma}=0,
\]
where $\vert$ denotes covariant differentiation with respect to the
Christoffel symbol. We now have
\[
u_\lambda\biggl(\frac{du^\lambda}{d\tau}
+\left\{{\lambda\atop \alpha\beta}\right\}^{(0)}u^\alpha u^\beta\biggr)
=u_\lambda {u^\lambda}_{\vert\mu}u^\mu=0.
\]
Multiplying (\ref{motion}) by $u_\mu$, we get
\[
\frac{dm_t}{d\tau}=m_t u_\mu {\tilde f}^\mu.
\]
We now obtain the equation of motion of the test particle (we drop the
$(0)$
notation):
\begin{equation}
\label{equationmotion}
\frac{du^\mu}{d\tau}+\left\{{\mu\atop \alpha\beta}\right\}u^\alpha
u^\beta
={\tilde f}^\mu-u_\alpha {\tilde f}^\alpha u^\mu.
\end{equation}
We see that, depending on the model chosen for $K^{[\mu\nu]}$, there
could
be a contribution due to the non-conservation of the mass, caused
by  an exchange of energy with the skew field $g_{[\mu\nu]}$.

According to Eq.(\ref{equationmotion}), test particles fall in an NGT
gravitational field
independently of their composition,
so the weak equivalence principle is satisfied in the new version of NGT.
However,
the strong equivalence principle is not satisfied in the theory, because
the non-gravitational laws of physics are not the same in different locally
Minkowskian
frames of reference; the skew part of the connection
$\Gamma^\lambda_{[\mu\nu]}$
is a tensor which cannot be transformed away at a point, in contrast to the
Christoffel
connection $\left\{{\lambda\atop \mu\nu}\right\}$.

If we assume that $K^{[\mu\nu]}$ is small and can be neglected, we
obtain the
geodesic equation of motion for test particles\cite{LegareMoff2}:
\begin{equation}
\label{geodesic}
\frac{du^\mu}{d\tau}+\left\{{\mu\atop \alpha\beta}\right\}u^\alpha
u^\beta=0.
\end{equation}

\section{The Field Equations for a Spherically Symmetric System}

For the case of a spherically symmetric field, the canonical form of
$g_{\mu\nu}$ in NGT is given by
\[
g_{\mu\nu}=\left(\matrix{-\alpha&0&0&w\cr
0&-\beta&f\hbox{sin}\theta&0\cr 0&-f\hbox{sin}\theta&
-\beta\hbox{sin}^2
\theta&0\cr-w&0&0&\gamma\cr}\right),
\]
where $\alpha,\beta,\gamma$ and $w$ are functions of $r$ and $t$. The
tensor $g^{\mu\nu}$ has the components:
\[
g^{\mu\nu}=\left(\matrix{{\gamma\over w^2-
\alpha\gamma}&0&0&{w\over w^2-\alpha\gamma}\cr
0&-{\beta\over \beta^2+f^2}&{f\hbox{csc}\theta\over
\beta^2+f^2}&0\cr
0&-{f\hbox{csc}\theta\over
\beta^2+f^2}&-{\beta\hbox{csc}^2\theta\over
\beta^2+f^2}&0\cr-{w\over w^2-\alpha\gamma}&0&0&-{\alpha\over
w^2-\alpha\gamma}\cr}\right).
\]
We shall assume that
$w=0$ and only the $g_{[23]}$ component of $g_{[\mu\nu]}$ is
different from zero.
It can be
proved that only the static solution for $g_{[23]}$ satisfies the physical,
asymptotically flat boundary conditions\cite{Clayton,Cornish1}.

We have
\[
\sqrt{-g}=\hbox{sin}\theta[(\alpha\gamma-w^2)(\beta^2+f^2)]^{1/2}.
\]

The vector $W_\mu$ can be determined from:
\begin{equation}
\label{Wequation}
W_\mu=-{2\over \sqrt{-g}}s_{\mu\rho}{{\bf
g}^{[\rho\sigma]}}_{,\sigma}.
\end{equation}
For the spherically symmetric field with
$w=0$ it follows from (\ref{Wequation}) that $W_\mu=0$.
The field equations (\ref{NGTfieldequations}) and (\ref{gskewequation})
for the spherically symmetric system now take the simpler form:
\begin{eqnarray}
\label{SSfieldequations}
G_{\mu\nu}(\Gamma)+\Lambda
g_{\mu\nu}+\frac{1}{4}\mu^2C_{\mu\nu}&=&8\pi
T_{\mu\nu},\\
{{\bf g}^{[\mu\nu]}}_{,\nu}&=&0.
\end{eqnarray}

Let us now reexpress the conservation laws in a form suitable for
calculations
in a spherically symmetric system.
Using the Einstein notation, the compatibility equations are given by
\begin{equation}
{{\bf g}^{\mu+\nu-}}_{;\sigma}\equiv {{\bf g}^{\mu\nu}}_{,\sigma}
+{\bf g}^{\rho\nu}\Gamma^\mu_{\rho\sigma}
+{\bf g}^{\mu\rho}\Gamma^\nu_{\sigma\rho}
-{\bf g}^{\mu\nu}\Gamma^\alpha_{(\sigma\alpha)}=0,
\end{equation}
where
\[
{A^{\mu+}}_{;\sigma}={A^\mu}_{,\sigma}+A^\rho\Gamma^\mu_{\rho\sigma},
\]
and
\[
{A^{\mu-}}_{;\sigma}={A^\mu}_{,\sigma}+A^\rho\Gamma^\mu_{\sigma\rho}.
\]
We have
\begin{equation}
\hbox{Re}[(G_{\rho-\nu-}(\Gamma){\bf g}^{\sigma+\nu-})_{;\sigma}]
=\frac{1}{2}[(G_{\rho-\nu-}(\Gamma){\bf
g}^{\sigma+\nu-})_{;\sigma}
+(G_{\nu+\rho+}(\Gamma){\bf g}^{\nu+\sigma-})_{;\sigma}]=0.
\end{equation}
This can be written as
\begin{equation}
\hbox{Re}[(G_{\rho-\nu-}(\Gamma){\bf g}^{\sigma+\nu-})_{;\sigma}]
=\hbox{Re}[(G_{\rho\nu}(\Gamma){\bf g}^{\sigma\nu})_{,\sigma}
-\frac{1}{2}G_{\tau\nu}(\Gamma)({\bf
g}^{\sigma\nu}\Gamma^\tau_{\sigma\rho}
+{\bf g}^{\tau\sigma}\Gamma^\nu_{\rho\sigma})]=0,
\end{equation}
which leads to the result:
\begin{equation}
\hbox{Re}[(G_{\rho-\nu-}(\Gamma){\bf g}^{\sigma+\nu-})_{;\sigma}]
=\frac{1}{2}(G_{\rho\nu}(\Gamma){\bf g}^{\sigma\nu})_{,\sigma}
+\frac{1}{2}(G_{\nu\rho}(\Gamma){\bf g}^{\nu\sigma})_{,\sigma}
+\frac{1}{2}{\bf G}_{\tau\nu}(\Gamma){g^{\tau\nu}}_{,\rho}=0.
\end{equation}
This is the same as the Bianchi identities (\ref{Bianchi}).
{}From (\ref{SSfieldequations}), we have for the spherically symmetric
conservation laws:
\begin{equation}
\label{SSconserve}
\hbox{Re}[(T_{\rho-\nu-}{\bf g}^{\sigma+\nu-})_{;\sigma}]=0.
\end{equation}

For a comoving coordinate system, we obtain
\begin{equation}
\label{comovingvelocity}
u^0=\frac{1}{\sqrt{\gamma}},\quad u^r=u^{\theta}=u^{\phi}=0.
\end{equation}
We define
\begin{equation}
T_{\mu\nu}=g_{\mu\beta}g_{\alpha\nu}T^{\alpha\beta},
\end{equation}
which is a Hermitian symmetric tensor, $T_{\mu\nu}={\tilde
T}_{\nu\mu}$, when
$g_{\mu\nu}$
is defined to be Hermitian symmetric: $g_{\mu\nu}={\tilde
g}_{\nu\mu}$.
{}From (\ref{contraction}) and (\ref{eq:energytensor}) we get
\[
T=\rho-3p+g_{[\alpha\beta]}K^{[\alpha\beta]}
=\rho - 3p +2fK,
\]
where we have defined: $K^{[23]}=K/\sin\theta$.

The NGT field equations in the presence of sources are given by
\begin{mathletters}
\begin{eqnarray}
R_{11}(\Gamma)&=&-{1\over 2}A^{''}-{1\over 8}[(A^\prime)^2+4B^2]
+{\alpha^\prime A^\prime\over 4\alpha}
+{\gamma^\prime\over 2\gamma}\biggl({\alpha^\prime\over 2\alpha}
-{\gamma^\prime\over 2\gamma}\biggr) \nonumber \\
& & \mbox{}
-\biggl({\gamma^\prime\over 2\gamma}\biggr)^\prime
+{\partial\over \partial t}\biggl({{\dot \alpha}\over 2\gamma}\biggr)
+{{\dot \alpha}\over 2\gamma}\biggl({{\dot\gamma}\over 2\gamma}
-{{\dot\alpha}\over 2\alpha}
+{1\over 2}{\dot A}\biggr)\nonumber \\
&& \mbox{}
+\Lambda\alpha-\frac{1}{4}\mu^2\frac{\alpha
f^2}{\beta^2+f^2}\nonumber \\
&=&4\pi\alpha(\rho-p+2fK), \\
R_{22}(\Gamma)=
R_{33}(\Gamma)\hbox{cosec}^2\theta&=&1+
\biggl({2fB-\beta A^\prime\over 4\alpha}\biggr)^\prime
+\biggl({2fB-\beta A^\prime\over
8\alpha^2\gamma}\biggr)(\alpha^\prime\gamma
+\gamma^\prime\alpha) \nonumber \\
& & \mbox{}
+{B(fA^\prime+2\beta B)\over 4\alpha}
-{\partial\over \partial t}\biggl({2fD-\beta{\dot A}\over
4\gamma}\biggr)
-{2fD-\beta{\dot A}\over 8\alpha\gamma^2}({\dot\alpha}\gamma
+{\dot \gamma}{\alpha}) \nonumber \\
& &  \mbox{}
-{D\over 4\gamma}(f{\dot A}+2\beta D)
+\Lambda\beta+\frac{1}{4}\mu^2\frac{\beta
f^2}{\beta^2+f^2}\nonumber \\
&=&4\pi\beta(\rho-p +2fK), \\
R_{00}(\Gamma)
&=&-{1\over 2}{\ddot A}-{1\over 8}({\dot
A}^2+4D^2)+{{\dot\gamma}\over
4\gamma}{\dot A}+{{\dot\alpha}\over
2\alpha}\biggl({{\dot\gamma}\over 2\gamma}
-{{\dot\alpha}\over 2\alpha}\biggr) \nonumber \\
& & \mbox{}
-{\partial\over \partial t}\biggl({{\dot\alpha}\over 2\alpha}\biggr)
+\biggl({\gamma^\prime\over 2\alpha}\biggr)^\prime
+{\gamma^\prime\over 2\alpha}\biggl({\alpha^\prime\over 2\alpha}
-{\gamma^\prime\over 2\gamma}+{1\over
2}A^\prime\biggr)\nonumber\\
&&\mbox{}
-\Lambda\gamma+\frac{1}{4}\mu^2\frac{\gamma f^2}
{\beta^2+f^2}\nonumber \\
&=&4\pi\gamma(\rho+3p - 2fK), \\
R_{[10]}(\Gamma)&=&0, \\
R_{(10)}(\Gamma)&=& -\frac{1}{2}{\dot A}'+\frac{1}{4}A'
\biggl(\frac{{\dot\alpha}}{\alpha}-\frac{1}{2}{\dot A}\biggr)
+\frac{1}{4}\frac{\gamma' {\dot
A}}{\gamma}-\frac{1}{2}BD\nonumber \\
&=&0,\\
R_{[23]}(\Gamma)&=&\sin\theta\biggl[\biggl({fA^\prime+2\beta B\over
4\alpha}\biggr)^\prime+{1\over 8\alpha}(fA^\prime+2\beta B)
\biggl({\alpha^\prime\over \alpha}+{\gamma^\prime\over\gamma}\biggr)
\nonumber
\\
& & \mbox{}
-{B\over 4\alpha}(2fB-\beta A^\prime)-{1\over 8\gamma}(f{\dot
A}+2\beta D)
\biggl({{\dot\gamma}\over \gamma}+{{\dot\alpha}\over
\alpha}\biggr)\nonumber \\
& & \mbox{}
-{\partial\over \partial t}\biggl({f{\dot A}+2\beta D\over
4\gamma}\biggr)
+{D\over 4\gamma}(2fD-\beta{\dot A})\biggr]\nonumber\\
&&\mbox{}-\biggl[\Lambda f-\frac{1}{4}\mu^2 f
\biggl(1+\frac{\beta^2}{\beta^2+f^2}\biggr)\biggr]\sin\theta\nonumber\\
&=& - 4\pi f\sin\theta(\rho - p).
\end{eqnarray}
\end{mathletters}
Here, prime denotes differentiation with respect to $r$, ${\dot A}
=\partial A/\partial t$, and we
have used the notation:
\begin{mathletters}
\begin{eqnarray}
\label{Aequation}
A&=&\hbox{ln}(\beta^2+f^2), \\
\label{Bequation}
B&=&{f\beta^\prime-\beta f^\prime\over
\beta^2+f^2}, \\
\label{Dequation}
D&=&{{\dot\beta}f-{\dot f}\beta\over \beta^2+f^2}.
\end{eqnarray}
\end{mathletters}
\section{Static Equilibrium Equations}

Let us consider the static hydrodynamic equations for stellar equilibrium.
For a static spherically symmetric system $g_{\mu\nu}, p, \rho$ and
$K^{[\mu\nu]}$ are functions only of the radial coordinate $r$.
For a fluid at rest, Eq.(\ref{comovingvelocity}) determines the velocity
components.  To simplify the field equations, let us assume that the direct
coupling
of
$g_{[23]}$ to $K^{[23]}$ is small and can be neglected. We also adopt
the
long-range approximation $\mu\approx 0$ and we choose $\Lambda=0$.
The field equations are given by
\begin{mathletters}
\begin{eqnarray}
R_{11}(\Gamma)&=&-\frac{1}{2}A''-\frac{1}{8}[(A')^2+4B^2]
+\frac{\alpha' A'}{4\alpha}+\frac{\gamma'}{2\gamma}
\biggl(\frac{\alpha'}{2\alpha}-\frac{\gamma'}{2\gamma}\biggr)\nonumber \\
&=&4\pi\alpha(\rho-p),\\
R_{22}(\Gamma)=R_{33}(\Gamma)\hbox{cosec}^2\theta
&=&1+\biggl(\frac{2fB-\beta A'}{4\alpha}\biggr)'
+\biggl(\frac{2fB-\beta
A'}{8\alpha^2\gamma}\biggr)(\alpha'\gamma+\gamma'\alpha)\nonumber\\
&&\mbox{}
+\frac{B(fA'+2\beta B)}{4\alpha}\nonumber\\
&=&4\pi\beta(\rho-p),\\
R_{00}(\Gamma)&=&\biggl(\frac{\gamma'}{2\alpha}\biggr)'+\frac{\gamma'}{2\alpha}
\biggl(\frac{\alpha'}{2\alpha}-\frac{\gamma'}{2\gamma}
+\frac{1}{2}A'\biggr)\nonumber\\
&=&4\pi\gamma(\rho+3p),\\
R_{[10]}(\Gamma)&=&0,\\
R_{(10)}(\Gamma)&=&0,\\
R_{[23]}(\Gamma)&=&\sin\theta
\biggl[\biggl(\frac{fA'+2\beta B}{4\alpha}\biggr)'
+\frac{1}{8\alpha}(fA'+2\beta B)\biggl(\frac{\alpha'}{\alpha}
+\frac{\gamma'}{\gamma}\biggr)
-\frac{B}{4\alpha}(2fB-\beta A')\biggr]\nonumber\\
&=&-4\pi f\sin\theta(\rho-p).
\end{eqnarray}
\end{mathletters}

The conservation laws (\ref{SSconserve}) give
\begin{equation}
\label{pconserve}
p'=-\frac{1}{2}(\rho+p)(\ln\gamma)'.
\end{equation}

We can further simplify the equations by adopting two approximation
schemes:
\begin{equation}
 r^2 \gg f(r),
\end{equation}
and
\begin{equation}
f(r) \gg r^2,
\end{equation}
where we have chosen $\beta(r)=r^2$. For the first approximation
scheme, we obtain for the $11/\alpha, 22/\beta, 00/\gamma$ and $[23]/f$ field
equations:
\begin{mathletters}
\begin{eqnarray}
-\frac{\gamma^{\prime\prime}}{2\alpha\gamma}
+\frac{\gamma'}{4\alpha\gamma}\biggl(\frac{\alpha'}{\alpha}
+\frac{\gamma'}{\gamma}\biggr)+\frac{\alpha'}{\alpha^2r}+I&=&4\pi(\rho-p),\\
\label{22equation}
\frac{1}{r^2}-\frac{1}{\alpha r^2}
+\frac{1}{2r\alpha}\biggl(\frac{\alpha'}{\alpha}
-\frac{\gamma'}{\gamma}\biggr)+N&=&4\pi(\rho-p),\\
\frac{\gamma^{\prime\prime}}{2\alpha\gamma}
-\frac{\gamma'}{4\alpha\gamma}\biggl(\frac{\alpha'}{\alpha}
+\frac{\gamma'}{\gamma}\biggr)+\frac{\gamma'}{r\alpha\gamma}
+L&=&4\pi(\rho+3p),\\
-\frac{\alpha'f'}{4\alpha^2f}+\frac{\gamma'f'}{4\alpha\gamma f}
-\frac{\gamma'}{r\alpha\gamma}+\frac{f^{\prime\prime}}{2\alpha f}
-\frac{f^\prime}{r\alpha f}+\frac{\alpha'}{r\alpha^2}
&=&4\pi(\rho-p).
\end{eqnarray}
\end{mathletters}
Here, we have defined
\begin{mathletters}
\begin{eqnarray}
I(r)&=&-\frac{\alpha'f^2}{\alpha^2r^5}+\frac{8ff'}{\alpha r^5}-\frac{8f^2}
{\alpha r^6}-\frac{3f'^2}{2\alpha r^4}+\frac{\alpha'ff'}{2\alpha^2 r^4}
-\frac{ff^{\prime\prime}}{\alpha r^4},\\
N(r)&=&-\frac{\alpha'f^2}{\alpha^2 r^5}-\frac{2f^2}{\alpha r^6}
+\frac{\gamma'f^2}{\alpha\gamma r^5}+\frac{\alpha'ff'}{2\alpha^2 r^4}
-\frac{ff"}{\alpha r^4}+\frac{3ff'}{\alpha r^5}-\frac{f'^2}{2\alpha r^4}
-\frac{\gamma' ff'}{2\alpha\gamma r^4},\\
L(r)&=&\frac{\gamma'ff'}{2\alpha\gamma r^4}-\frac{\gamma'f^2}
{\alpha\gamma r^5}.
\end{eqnarray}
\end{mathletters}

Let us now consider the following combination:
$11/2\alpha+22/\beta+00/2\gamma$:
\begin{equation}
\label{00equation}
\frac{\alpha'}{r\alpha^2}+\frac{1}{r^2}-\frac{1}{\alpha r^2}+G=8\pi\rho,
\end{equation}
where
\[
G(r)=\frac{1}{2}[I(r)+L(r)]+N(r).
\]
Eq.(\ref{00equation}) can be written:
\begin{equation}
\biggl(\frac{r}{\alpha}\biggr)'=1-8\pi(\rho-\frac{G}{8\pi})r^2.
\end{equation}
Integrating this equation gives
\begin{equation}
\label{alphaequation}
\alpha(r)=\frac{1}{1-2{\cal M}(r)},
\end{equation}
where
\begin{equation}
{\cal M}(r)=4\pi\int_0^r dr'r^{'2}{\tilde\rho}(r'),
\end{equation}
and
\begin{equation}
\label{tilderho}
{\tilde\rho}(r)=\rho(r)-\frac{G(r)}{8\pi}.
\end{equation}

{}From (\ref{pconserve}), we have
\begin{equation}
\label{ratioeq}
\frac{\gamma'}{\gamma}=-\frac{2p'}{p+\rho}.
\end{equation}
By using Eqs.(\ref{22equation}), (\ref{alphaequation}) and (\ref{ratioeq}),
we obtain the extended Oppenheimer-Volkoff
equation\cite{OppyVolkoff,Weinberg}:
\begin{equation}
\label{extoppyvolkoff}
p'(r)=-\frac{{\cal M}(r){\rho}(r)}{r^2}
\biggl(1+\frac{p(r)}{\rho(r)}\biggr)
\biggl(1+\frac{4\pi r^3 P(r)}{{\cal M}(r)}\biggr)
\biggl(1-\frac{2{\cal M}(r)}{r}\biggr)^{-1},
\end{equation}
where
\begin{equation}
\label{Pequation}
P(r)=4\pi\biggl[p(r)+\biggl(\frac{N(r)-G(r)}{4\pi}\biggr)\biggr].
\end{equation}

Eq.(\ref{extoppyvolkoff}) reduces in the non-relativistic limit to the
Newtonian
equation for stellar hydrodstatic equilibrium. It is the same as the
Oppenheimer-Volkoff equation, except that $\rho(r)$ is replaced by
${\tilde \rho}(r)$ in ${\cal M}(r)$ and $p(r)$ is replaced by
$P(r)$ in the third factor on the right-hand side.

By using (\ref{extoppyvolkoff}) we get
\begin{equation}
\frac{\gamma'(r)}{\gamma(r)}=
\frac{2}{r^2}[{\cal M}(r)+4\pi r^3 P(r)]
\biggl(1-\frac{2{\cal M}(r)}{r}\biggr)^{-1}.
\end{equation}
The solution that has $\gamma(\infty)=1$ is given by
\begin{equation}
\gamma(r)=\exp\biggl\{-2\int^\infty_r\frac{dr'}{r'^2}
[{\cal M}(r')+4\pi r^{'3}P(r')]
\biggl[1-\frac{2{\cal M}(r')}{r'}\biggr]\biggr\}.
\end{equation}
These approximate results hold for a weak $f(r)$ field and for moderately
relativistic systems such as neutron stars.

Consider now the second approximation scheme that holds for large
$f(r)$ with $f(r) \gg  r^2$. We get for the $11/\alpha, 22/\beta, 00/\gamma$
and
$[23]/f$ field equations:
\begin{mathletters}
\begin{eqnarray}
\frac{f'^2}{2\alpha f^2}-\frac{f^{\prime\prime}}
{\alpha f}+\frac{\alpha'f'}{2f\alpha^2}
+\frac{\gamma'^2}{4\alpha\gamma^2}+\frac{\gamma'\alpha'}{4\gamma\alpha^2}
-\frac{\gamma^{\prime\prime}}{2\alpha\gamma}-\frac{5f'^2r^4}{2\alpha
f^4}\nonumber\\
+\frac{f^{\prime\prime}r^4}
{\alpha f^3}-\frac{\alpha'f'r^4}{2\alpha^2f^3}+\frac{\alpha'r^3}
{\alpha^2 f^2}+\frac{8f'r^3}{\alpha f^3}-\frac{8r^2}{\alpha f^2}
&=&4\pi(\rho-p),\\
\frac{1}{r^2}-\frac{f'}{\alpha fr}-\frac{\gamma'f'}{2\alpha\gamma f}
-\frac{\alpha'}{2\alpha^2r}-\frac{f^{\prime\prime}}{\alpha f}
+\frac{\alpha'f'}{2\alpha^2f}+\frac{1}{\alpha r^2}+\frac{\gamma'}
{2\alpha\gamma r}+\frac{f'^2}{2\alpha f^2}
&=&4\pi(\rho-p),\\
-\frac{\gamma'\alpha'}{4\alpha^2\gamma}+\frac{\gamma'f'}{2\alpha\gamma f}
+\frac{\gamma^{\prime\prime}}{2\alpha\gamma}
-\frac{\gamma'^2}{4\alpha\gamma^2}
+\frac{\gamma'r^3}{\alpha\gamma f^2}-\frac{\gamma'f' r^4}{2\alpha\gamma f^3}
&=&4\pi(\rho+3p),\\
\label{fstrongeq}
-\frac{\alpha'f' r^4}{2\alpha^2 f^3}-\frac{f'^2r^4}{\alpha f^4}
-\frac{f^{\prime\prime}}{2\alpha f}
+\frac{\alpha'f'}{4\alpha^2f}-\frac{\gamma'f'}{4\alpha\gamma f}
+\frac{3f'r^3}{\alpha f^3}-\frac{\gamma' r^3}{\alpha\gamma f^2}
-\frac{4r^2}{\alpha f^2}\nonumber\\
+\frac{\alpha'r^3}{\alpha^2f^2}+\frac{f^{\prime\prime}r^4}{\alpha f^3}
+\frac{\gamma'f'r^4}{2\alpha\gamma f^3}
&=&4\pi(\rho-p).
\end{eqnarray}
\end{mathletters}

By forming the combination: $11/\alpha+00/\gamma$, we get
\begin{eqnarray}
\frac{f'^2}{2\alpha f^2}-\frac{f^{\prime\prime}}{\alpha
f}+\frac{\alpha'f'}{2\alpha^2f}
-\frac{5f'^2r^4}{2\alpha f^4}+\frac{f^{\prime\prime}r^4}{\alpha f^3}
-\frac{\alpha'f'r^4}{2\alpha^2f^3}+\frac{8f'r^3}{\alpha f^3}
-\frac{8r^2}{\alpha f^2}\nonumber\\
+\frac{\alpha' r^3}{\alpha^2 f^2}+\frac{\gamma'}{\gamma}
\biggl(\frac{f'}{2\alpha f}+\frac{r^3}{\alpha f^2}-\frac{f'r^4}{2\alpha
f^3}\biggr)
&=&8\pi(\rho-p).
\end{eqnarray}
Using (\ref{ratioeq}), we find the hydrostatic equilibrium equation:
\begin{eqnarray}
p'=-\biggl[\frac{\alpha f^3(\rho+p)}{(f'f^2+2fr^3-f'r^4)}\biggr]
\biggl[\frac{f^{\prime\prime}}{\alpha f}-\frac{f'^2}{2\alpha
f^2}-\frac{\alpha'f'}{2\alpha^2f}+\frac{5f'^2r^4}{2\alpha f^4}
-\frac{f^{\prime\prime}r^4}{\alpha f^3}\nonumber\\
+\frac{\alpha'f'r^4}{2\alpha^2f^3}-\frac{8f'r^3}{\alpha f^3}
+\frac{8r^2}{\alpha f^2}-\frac{\alpha'r^3}{\alpha^2f^2}+8\pi(\rho-p)\biggr].
\end{eqnarray}

Near $r=0$, we get
\begin{equation}
\label{strongfeq}
p'=-\biggl[\frac{\alpha
f(\rho+p)}{f'}\biggr]\biggl[\frac{f^{\prime\prime}}{\alpha f}
-\frac{f'^2}{2\alpha f^2}-\frac{\alpha'f'}{2\alpha^2f}
+8\pi(\rho-p)\biggr].
\end{equation}

We adopt the equation of state for a gas of highly relativistic particles:
\begin{equation}
\label{relativisticgas}
p=\frac{1}{3}\rho.
\end{equation}
Using (\ref{relativisticgas}) in (\ref{strongfeq}) gives
\begin{equation}
\rho'= -\biggl(\frac{4\alpha f\rho}{f'}\biggr)
\biggl(\frac{f^{\prime\prime}}{\alpha f}
-\frac{f'^2}{2\alpha f^2}
-\frac{\alpha'f'}{2\alpha^2f}+\frac{16\pi}{3}\rho\biggr)
\end{equation}
In the limit of large densities, $\rho\rightarrow\infty$, we get
\begin{equation}
\rho'=-\biggl(\frac{64\pi}{3}\biggr)\biggl(\frac{\alpha f}{f'}\biggr)\rho^2.
\end{equation}
This equation has the formal solution:
\begin{equation}
\rho(r)=\exp\biggl[-\frac{64\pi}{3}\int^rdr'\frac{\alpha(r')f(r')\rho(r')}{f'(r')}\biggr]
+ \hbox{const}.
\end{equation}

Thus, in the limit of  large $\rho$ and for $f/f' < 0$, at the center of a
dense collapsed object, only repulsive forces occur which increase with
increasing density. For either a compact object of high density or a very
massive object, we expect that an equilibrium state can be achieved.

\section{Spherically Symmetric Collapse Equations for Dust}

If we adopt the approximation scheme leading to the geodesic equation,
Eq.(\ref{geodesic}),
for falling test particles, then we can use a comoving coordinate system
with the velocity components:
\[
u^0=1,\quad u^r=u^\theta=u^\phi=0,
\]
and the time dependent metric in normal Gaussian form:
\begin{equation}
ds^2=dt^2-\alpha(r,t)dr^2-\beta(r,t)(d\theta^2+\hbox{sin}^2\theta
d\phi^2).
\end{equation}

In order to simplify the field equations, we must make several
approximations. We shall assume that the star collapses as a pressureless
dust with $p=0$ and that the direct coupling term in the source
tensor, $K_{[\mu\nu]}$, is small and can be neglected during the
collapse. As in the last section, we also set $\Lambda=0$ and take the
approximation $\mu\approx 0$. The field equations now take the form:
\begin{mathletters}
\label{eq:matteqs}
\begin{eqnarray}
\label{eq:matteqs_11}
-{1\over 2\alpha}A''-{1\over 8\alpha}[(A^\prime)^2+4B^2]
+{\alpha^\prime A^\prime\over 4\alpha^2}
+{\ddot {\alpha}\over 2\alpha}-\frac{{\dot{\alpha}}^2}{4\alpha^2}
+{\dot{\alpha}\dot{A}\over 4\alpha}&=&4\pi \rho , \\
\label{eq:matteqs_22}
{1\over\beta}+
{1\over 4\beta}\biggl({2fB-\beta A^\prime\over \alpha}\biggr)^\prime
+{\alpha^\prime\over 8\beta}
\biggl({2fB-\beta A^\prime\over \alpha^2}\biggr)\nonumber \\
 +{1\over 4}{B(fA^\prime+2\beta B)\over \alpha\beta}
+{1\over 4\beta}\biggl(\dot{\beta}\dot{A}+\beta\ddot{A}\biggr)
-{1\over 2\beta}({\dot{f}}D+f{\dot D})
+{\dot{A}\dot {\alpha}\over 8\alpha} \nonumber \\
-{1\over 4}\biggl({fD\dot\alpha\over \alpha\beta}\biggr)
-{D\over 4\beta}(f\dot{A}+2\beta D)&=&4\pi \rho, \\
\label{eq:matteqs_00}
-{1\over 2}{\ddot A}-{1\over 8}({\dot A}^2+4D^2)
+{\dot{\alpha}^2\over 4\alpha^2}
-{1\over 2}{\ddot{\alpha}\over \alpha}&=&4\pi\rho, \\
\label{eq:matteqs_(01)}
-{1\over 2}{\dot A}^\prime+{1\over 4}{A^\prime\dot\alpha\over \alpha}
-{1\over 8}A^\prime\dot{ A}-{1\over 2}BD&=&0, \\
\label{eq:matteqs_[23]}
-{1\over 4f}\biggl({fA^\prime+2\beta B\over \alpha}\biggr)^\prime
-{1\over 8}{(fA^\prime+2\beta B)\alpha^\prime\over  \alpha^2f}
+{1\over 4}{B(2fB-\beta A^\prime)\over \alpha f} \nonumber \\
+{1\over 8}{(f\dot{A}+2\beta D)\dot{\alpha}\over \alpha f}
+{1\over 4f}(\dot{f}\dot{A}+f\ddot{A}+2\dot{\beta}D+2\beta\dot{D})
-{D\over 4f}(2fD-\beta\dot{A})&=&4\pi\rho.
\end{eqnarray}
\end{mathletters}

{}From the conservation law (\ref{SSconserve}), we obtain
within our approximation scheme:
\[
\dot{\rho}+ \rho\biggl({{\dot \alpha}\over 2\alpha}
+{{\dot \beta}\over \beta}\biggr)=0.
\]
{}From this result, it follows that
\begin{equation}
\label{conservationlaw}
{\partial\over \partial t}(\rho\beta\sqrt{\alpha})=0.
\end{equation}

There are two approximate regimes that we can adopt in order to further
simplify the
set of field
equations. In the first one, it is assumed that
\begin{equation}
\label{approx1}
\beta(r,t) \gg f(r,t),
\end{equation}
while in the second one, we have
\begin{equation}
\label{approx2}
f(r,t) \gg \beta(r,t).
\end{equation}

In the first approximation scheme using (\ref{approx1}), we obtain from
(\ref{eq:matteqs_11})--(\ref{eq:matteqs_[23]}) the equations:
\begin{mathletters}
\begin{eqnarray}
\label{eq:simpleeqs_11}
-{1\over \alpha}\biggl[{\beta'' \over \beta}-{\beta^{' 2}\over 2\beta^2}
-{\alpha^\prime\beta^\prime\over
2\alpha\beta}\biggr]+{\ddot{\alpha}\over 2\alpha}
-{\dot\alpha^2\over {4\alpha^2}}+{\dot\alpha\dot\beta\over 2\alpha\beta}
+W&=&4\pi\rho, \\
\label{eq:simpleeqs_22}
{1\over \beta}-{1\over \alpha}\biggl({\beta'' \over 2\beta}
-{\alpha^\prime\beta^\prime\over 4\alpha\beta}\biggr)
+{{\ddot{\beta}}\over 2\beta}+{{\dot{\alpha}}{\dot{\beta}}\over
4\alpha\beta}+ X
&=&4\pi\rho, \\
\label{eq:simpleeqs_00}
-{{\ddot{\alpha}}\over 2\alpha}-{{\ddot{\beta}}\over
\beta}+{{\dot{\alpha}}^2\over
4\alpha^2}+{{\dot{\beta}}^2\over 2\beta^2}+ Y&=&4\pi\rho, \\
\label{eq:simpleeqs_(01)}
-{{\dot{\beta}}^\prime\over \beta}+{\beta^\prime{\dot{\beta}}\over
2\beta^2}+{{\dot{\alpha}}\beta^\prime\over 2\alpha\beta}+ Z&=&0,\\
\label{eq:simpleeqs_[23]}
\frac{\dot f\dot\beta}{2\beta f}-\frac{\ddot
f}{2f}+\frac{\beta'^2}{2\alpha\beta^2}
-\frac{{\dot\beta^2}}{2\beta^2}+\frac{\alpha'\beta'}{2\alpha^2\beta}
-\frac{\beta^{\prime\prime}}{\alpha\beta}+\frac{\ddot\beta}{\beta}
-\frac{f'\beta'}{2\alpha\beta f}
\nonumber\\
-\frac{\dot\alpha\dot f}{4\alpha f}+\frac{\dot\alpha\dot\beta}{2\alpha\beta}
+\frac{f^{\prime\prime}}{2\alpha f}-\frac{\alpha' f'}{4\alpha^2 f}&=&4\pi\rho.
\end{eqnarray}
\end{mathletters}
Here, we have defined
\begin{mathletters}
\begin{eqnarray}
\label{Wexpression}
W(r,t)&=&-\frac{\alpha'\beta'f^2}{2\alpha^2\beta^3}
+\frac{\beta^{\prime\prime}f^2}
{\alpha\beta^3}-\frac{\dot\alpha\dot\beta f^2}{2\alpha^2\beta^3}
\nonumber\\
& &\mbox{}
-\frac{\dot\alpha\dot\beta f^2}{2\alpha\beta^3}-\frac{5\beta'^2f^2}
{2\alpha\beta^4}+\frac{\dot\alpha f\dot f}{2\alpha\beta^2}
+\frac{\alpha'ff'}{2\alpha^2\beta^2}-\frac{ff^{\prime\prime}}{\alpha\beta^2}
+\frac{4ff'\beta'}{\alpha\beta^3}-\frac{3f'^2}{2\alpha\beta^2},\\
X(r,t)&=&-\frac{\dot\alpha\dot\beta f^2}{2\alpha\beta^3}-\frac{\alpha'\beta'
f^2}{2\alpha^2\beta^3}+\frac{\beta^{\prime\prime}f^2}{\alpha\beta^3}
-\frac{\ddot\beta f^2}
{\beta^3}+\frac{\dot\alpha f\dot f}{2\alpha\beta^2}
+\frac{\dot\beta^2f^2}{\beta^4}
-\frac{ff^{\prime\prime}}{\beta^2\alpha}-\frac{\beta'^2f^2}{\alpha\beta^4}
\nonumber\\
& & \mbox{}
+\frac{\alpha'ff'}{2\alpha^2\beta^2}+\frac{\dot f^2}{2\beta^2}
+\frac{f\ddot f}{\beta^2}-\frac{f'^2}{2\alpha\beta^2}
-\frac{3f\dot f\dot\beta}{2\beta^3}+\frac{3ff'\beta'}{2\alpha\beta^3},\\
Y(r,t)&=&\frac{\ddot\beta f^2}{\beta^3}-\frac{5\dot\beta^2f^2}{2\beta^4}
-\frac{3\dot f^2}{2\beta^2}+\frac{4\dot\beta f\dot f}{\beta^3}
-\frac{f\ddot f}{\beta^2},\\
Z(r,t)&=&\frac{\dot\beta'f^2}{\beta^3}-\frac{5\dot\beta\beta' f^2}{2\beta^4}
-\frac{\dot\alpha\beta' f^2}{2\alpha\beta^3}+\frac{2\dot\beta ff'}{\beta^3}
-\frac{f\dot f'}{\beta^2}-\frac{3f'\dot f}{2\beta^2}\nonumber\\
& & \mbox{}
+\frac{\dot\alpha ff'}{2\alpha\beta^2}+\frac{2\beta'f\dot f}{\beta^3}.
\end{eqnarray}
\end{mathletters}

In the second approximation scheme, we have for the $11/\alpha,
22/\beta, 00, (01)$ and $[23]/f$ components:
\begin{mathletters}
\begin{eqnarray}
\label{comp11}
\frac{{\alpha'}{f'}}{2f\alpha^2}+\frac{f'^2}{2\alpha f^2}
-\frac{f^{\prime\prime}}{\alpha
f}+\frac{\ddot\alpha}{2\alpha}-\frac{{\dot\alpha}^2}{4\alpha^2}
+\frac{\dot\alpha\dot f}{2\alpha f}-\frac{\alpha'f'\beta^2}
{2\alpha^2 f^3}\nonumber\\
-\frac{5f'^2\beta^2}{2\alpha f^4}+\frac{f^{\prime\prime}\beta^2}{\alpha f^3}
-\frac{\dot\alpha\dot f\beta^2}{2\alpha f^3}+\frac{\dot\alpha\dot\beta\beta}
{2\alpha f^2}-\frac{\beta^{\prime\prime}\beta}{\alpha
f^2}-\frac{3{\beta'}^2}{2\alpha f^2}
+\frac{4f'\beta'\beta}{\alpha f^3}+\frac{\alpha'\beta'\beta}{2\alpha^2f^2}
&=&4\pi\rho,\\
\label{comp22}
\frac{1}{\beta}-\frac{\dot\alpha\dot\beta}{4\alpha\beta}-\frac{{\dot
f}^2}{2f^2}
+\frac{\ddot f}{f}+\frac{\dot\alpha\dot f}{2\alpha
f}+\frac{\alpha'f'}{2\alpha^2f}
-\frac{\ddot\beta}{2\beta}-\frac{f^{\prime\prime}}{\alpha f}+\frac{\dot f
\dot\beta}
{2\beta f}\nonumber\\
+\frac{\beta^{\prime\prime}}{2\alpha\beta}-\frac{\alpha'\beta'}{4\alpha^2\beta}
-\frac{\beta'f'}{2\alpha\beta f}+\frac{f'^2}{2\alpha f^2}&=&4\pi\rho,\\
\label{comp00}
-\frac{\ddot\alpha}{2\alpha}+\frac{\dot{\alpha}^2}{4\alpha^2}+\frac{{\dot f}^2}
{2f^2}-\frac{\ddot f}{f}-\frac{5{\dot f}^2\beta^2}{2f^4}+\frac{{\ddot
f}\beta^2}
{f^3}-\frac{{\ddot\beta}\beta}{f^2}-\frac{3{\dot\beta}^2}{2f^2}
+\frac{4{\dot f}{\dot\beta}\beta}{f^3}&=&4\pi\rho,\\
\label{comp(01)}
\frac{f'\dot f}{2f^2}-\frac{\dot f'}{f}+\frac{\dot\alpha f'}{2\alpha f}
-\frac{5f'\dot f\beta^2}{2f^4}+\frac{\dot f'\beta^2}{f^3}
-\frac{\dot\alpha f'\beta^2}{2\alpha f^3}+\frac{2\dot
f\beta'\beta}{f^3}\nonumber\\
-\frac{3\dot\beta\beta'}{2f^2}+\frac{2\dot\beta\beta f'}{f^3}
-\frac{\dot\beta'\beta}{f^2}+\frac{\dot\alpha\beta'\beta}{2\alpha f^2}&=&0,\\
\label{comp[23]}
-\frac{3\dot f\dot\beta\beta}{2f^3}+\frac{\dot\beta^2}{2f^2}
-\frac{\dot\alpha\dot f\beta^2}{2\alpha
f^3}+\frac{3f'\beta'\beta}{2\alpha^2f^3}
-\frac{{f'}^2\beta^2}{\alpha f^4}-\frac{\beta'^2}{2\alpha f^2}
+\frac{\alpha'\beta'\beta}{2\alpha^2f^2}\nonumber\\
+\frac{\ddot f}{2f}-\frac{f^{\prime\prime}}{2\alpha f}
-\frac{\alpha' f'\beta^2}{2\alpha^2 f^3}
-\frac{\ddot f\beta^2}{f^3}+\frac{\dot\alpha\dot f}{4\alpha f}
+\frac{f^{\prime\prime}\beta^2}{\alpha f^3}
+\frac{{\dot f}^2\beta^2}{f^4}+\frac{\alpha' f'}
{4\alpha^2f}+\frac{\dot\alpha\dot\beta\beta}{2\alpha f^2}
-\frac{\beta^{\prime\prime}\beta}
{\alpha f^2}+\frac{{\ddot\beta}\beta}{f^2}&=&4\pi\rho.
\end{eqnarray}
\end{mathletters}
Here, we have used
\begin{equation}
B\approx\frac{f\beta'-\beta f'}{f^2},\quad D\approx
\frac{{\dot\beta}f-{\dot f}\beta}{f^2}.
\end{equation}
\section{Analysis of Dust Collapse}

We shall simplify our model for collapse even further and assume that the
density $\rho$ is independent of position. To analyze the field equations,
we shall follow the procedures given by Weinberg, Landau and Lifshitz, and by
Misner, Thorne and Wheeler\cite{Weinberg,Landau,MTW}.  Consider first the
approximation scheme determined by the condition (\ref{approx1}). It is
assumed that a solution can be found by a separation of variables:
\begin{equation}
\label{separationeq}
\alpha(r,t)=h(r)R^2(t),\quad \beta(r,t)=r^2S^2(t).
\end{equation}
{}From Eq.(\ref{eq:simpleeqs_(01)}), we get
\begin{equation}
\label{RSequation}
\frac{{\dot R}}{R}-\frac{{\dot S}}{S}=\frac{1}{2}Z(r,t)r.
\end{equation}
If, during the collapse, we assume that $Z(r,t)\approx 0$, then from
(\ref{RSequation}) we find that
\[
R(t)\approx S(t).
\]
Eqs. (\ref{eq:simpleeqs_11}) and (\ref{eq:simpleeqs_22}) now become
\begin{eqnarray}
\label{equation1}
{h^\prime(r)\over r h^2(r)}+{\ddot{R}}(t)R(t)+2{\dot{R}}^2(t)
+R^2(t)W(r,t)&=&4\pi R^2(t)\rho(t),\\
\label{equation2}
{1\over r^2}-{1\over r^2 h(r)}+{h^\prime(r)\over 2r
h^2(r)}+{\ddot{R}}(t)R(t)
+2{\dot{R}}^2(t)+R^2(t)X(r,t)&=&4\pi R^2(t)\rho(t).
\end{eqnarray}

The metric line-element takes the cosmological
Friedmann-Robertson-Walker
form:
\begin{equation}
\label{FRWmetric}
ds^2=dt^2-R^2(t)\biggl[h(r)dr^2+r^2(d\theta^2+\sin^2\theta
d\phi^2)\biggr].
\end{equation}
Thus, we assume that during the collapse the metric is approximately
isotropic and homogeneous.  If we assume that $W(r,t)\approx X(r,t)\approx 0$,
then we find from Eqs.(\ref{equation1}) and (\ref{equation2}) the GR solution:
\begin{equation}
\label{FRW}
h(r)={1\over 1-kr^2},
\end{equation}
where
\[
2k={h^\prime(r)\over r h^2(r)}={1\over r^2}-{1\over r^2 h(r)}
+{h^\prime(r)\over 2r h^2(r)},
\]
is a constant.

We observe at this point that by means of a Killing vector analysis, it can
be proved that a time dependent solution of NGT cannot describe an exact
homogeneous and isotropic spacetime, unless $g_{[\mu\nu]}$ is identically
zero\cite{Moffat4}.
Let us expand the metric $g_{(\mu\nu)}$ as
\begin{equation}
g_{(\mu\nu)}(r,t)=g^{HI}_{(\mu\nu)}(r,t)+\delta g_{(\mu\nu)}(r,t),
\end{equation}
where $g^{HI}_{(\mu\nu)}$ denotes the homogeneous and isotropic solution of
$g_{(\mu\nu)}$ and $\delta g_{(\mu\nu)}$ are small quantities which
break the maximally symmetric solution with constant Riemannian curvature.

Eqs.(\ref{eq:simpleeqs_00}) and (\ref{equation1}) can now be written:
\begin{eqnarray}
\label{eqn1}
2b(r)+{\ddot{R}}(t)R(t)+2{\dot{R}}^2(t)+R^2(t)W(r,t)&=&4\pi
R^2(t)\rho(t),\\
\label{eqn2}
-{\ddot R}(t)R(t)+\frac{1}{3}R^2(t)Y(t)&=&\frac{4\pi}{3}R^2(t)
\rho(t),
\end{eqnarray}
where
\[
2b(r)={h^\prime(r)\over r h^2(r)}.
\]
Eliminating ${\ddot R}$ by adding (\ref{eqn1}) and
(\ref{eqn2}), we get
\begin{equation}
\label{Rvelocityeq}
{\dot{R}}^2(t)=-b(r)+{8\pi\over 3}{\tilde\rho}(r,t)R^2(t),
\end{equation}
where
\begin{equation}
{\tilde\rho}(r,t)=\rho(t)-\frac{3}{8\pi}H(r,t)
\end{equation}
and
\[
H(r,t)=\frac{1}{2}[W(r,t)+\frac{1}{3}Y(t)].
\]

We shall normalize $R(t)$ so that
\[
R(0)=1,
\]
and we define
\[
{\tilde\rho}(r,t)={\tilde\rho}(r,0)R^{-3}(t).
\]
Let us assume that the fluid is at rest at $t=0$, so that ${\dot R}(0)=0$.
We have
\[
b(r)=\frac{8\pi}{3}{\tilde\rho}(r,0),
\]
where
\[
{\tilde\rho}(r,0)=\rho(0)-\frac{3}{8\pi}H(r,0).
\]
We now obtain the equation:
\begin{equation}
\label{velocityeq}
{\dot{R}}^2(t)=-b(r)+{8\pi\over 3}{\tilde\rho}(r,0)R^{-1}(t),
\end{equation}
which can be written as
\begin{equation}
\label{Reqn}
{\dot{R}}^2(t)=b(r)[R^{-1}(t)-1].
\end{equation}
This has the same form as the corresponding equation in GR,
except that $b(r)$ has the additional contribution from
the inhomogeneous, $r$ dependent quantity, $H(r,0)$, due to the skew
fields.

Eq.(\ref{Reqn}) has the parametric solution:
\begin{eqnarray}
t&=&\frac{\eta+\sin\eta}{2\sqrt{b}},\\
R&=&\frac{1}{2}(1+\cos\eta).
\end{eqnarray}
This solution reveals that $R(t)=0$ for $\eta=\pi$ and $t=t_0$ where
\[
t_0=\frac{\pi}{2\sqrt{b(r)}}
=\frac{\pi}{2}\biggl(\frac{3}{8\pi{\tilde\rho}(r,0)}\biggr)^{1/2}.
\]
Thus, as in GR, ${\tilde\rho}(r,t)\rightarrow\infty$ as $R(t)\rightarrow 0$
and the fluid sphere with initial density ${\tilde\rho}(r,0)>0$ and $p=0$ will
collapse
from rest to a state with infinite proper density in the finite time $t_0$,
provided that
${\tilde\rho}(r, t) > 0$. But ${\tilde\rho}(r,t)$ need not be
positive definite, for it contains second order skew curvature
contributions. If, for $R(t)$ near
zero, we have ${\tilde\rho}(r,t) < 0$, then the collapse could be stopped
even for the approximation $f(r,t) \ll \beta(r,t)$.

However, as we shall see in the following sections, the results for
$\beta(r,t)\gg f(r,t)$
are only expected to hold at the initial stage of the
collapse for $r \gg 2M$. As the collapse proceeds and the star becomes
more dense, we should use the second aproximate regime for which the condition
(\ref{approx2}) holds. As before, we assume that a separable solution is
possible:
\begin{equation}
\label{sepvariables}
\alpha(r,t)=q(r)R^2(t),\quad \beta(r,t)=r^2R^2(t).
\end{equation}
Then, equation (\ref{comp00}) becomes
\begin{eqnarray}
\label{fequation}
\frac{\ddot{R}}{R}&=&-4\pi\rho-\frac{\ddot{f}}{f}+\frac{{\dot f}^2}{2f^2}
-\frac{5{\dot f}^2r^4R^4}{2f^4}+\frac{\ddot f r^4R^4}{f^3}\nonumber\\
& & \mbox{}
-\frac{8r^4{\dot R}^2R^2}{f^2}-\frac{2r^4R^3{\ddot R}}{f^2}
+\frac{8{\dot f} r^4R^3{\dot R}}{f^3}.
\end{eqnarray}

We can learn about the behavior of NGT by considering the behavior of
(\ref{fequation}). Near $R(t)=0$, Eq.(\ref{fequation}) becomes
\begin{equation}
\frac{\ddot R}{R}=\frac{{\dot f}^2}{2f^2}-\frac{\ddot f}{f}-4\pi\rho.
\end{equation}
Let us assume that as the collapse approaches  $R(t)=0$:
\begin{equation}
\frac{{\dot f}^2}{2f^2}-\frac{\ddot f}{f}-4\pi\rho > 0.
\end{equation}
If for a contracting star we have $\dot{R}/R > 0$, and since by definition
$R >0$, then ${\ddot R} > 0$ and $R(t)$ will not pass through zero during the
collapse. Thus, the proper circumference and the proper three-volume of the
collapsing star remain finite at the end of the collapse. Moreover, the proper
density
$\rho(t)$ is finite for the final state of the star. In contrast, in GR we have
\begin{equation}
\frac{{\ddot R}}{R}=-\frac{4\pi}{3}\rho(0)R^{-3},
\end{equation}
from which it follows that the fluid sphere of initial density $\rho(0)>0$
and zero pressure must collapse from rest in a finite time $t_0$ to a state
of infinite proper energy density.

\section{Exterior Static Vacuum Solution}

The metric line element in NGT has the form
\begin{equation}
ds^2=m_{(\mu\nu)}dx^\mu dx^\nu,
\end{equation}
where $m_{(\mu\nu)}$ can have three forms in the long-range limit
$\mu\approx 0$:
\begin{mathletters}
\begin{eqnarray}
{}^1m_{(\mu\nu)}&=&g_{(\mu\nu)},\\
{}^2m^{(\mu\nu)}&=&g^{(\mu\nu)},\\
{}^3m^{(\mu\nu)}&=&\frac{2s}{g}(g^{(\mu\nu)})^{-1}-g^{(\mu\nu)},
\end{eqnarray}
\end{mathletters}
where $s=\hbox{Det}(g_{(\mu\nu)})$ and
$g=\hbox{Det}(g_{\mu\nu})$.
The metrics ${}^1m_{(\mu\nu)}, {}^2m^{(\mu\nu)}$
and ${}^3m^{(\mu\nu)}$ were obtained from a study of the Cauchy
evolution of field equations in Einstein's unified field theory by
Maurer-Tison\cite{Tison,Tonnelat}.
There are three light-cones in NGT, corresponding to the propagation of
different  zero mass modes. These three metrics can describe the causal
Minkowskian light-cone structure of the spacetime in NGT, although it is
expected
that when $\mu\not=0$, the light cone structure of the spacetime will be
somewhat
modified, since the massive $g_{[\mu\nu]}$ field will not propagate
causal information along the light cone. The three light cones degenerate to
the
standard single light cone of special relativity, when NGT is expanded about
the Minkowski background metric, and also when it is expanded about a GR
background metric. At the Schwarzschild radius, $r=2M$, the three light cones
are regrouped by interchanging their overlapping status\cite{Tison}.

In Born-Infeld non-linear electrodynamics\cite{BornInfeld}, there are two
kinds of electric fields, one of which is point-like and
singular at $r=0$, while the other one is
finite at $r=0$; for the latter the electric current density is spread out over
space. Similarly, for the spherically symmetric NGT vacuum solution,
the Riemannian geometries associated with the metrics ${}^1m, {}^2m$
and
${}^3m$ are singular at $r=0$, while the non-Riemannian geometry
determined by the
fundamental tensor
$g_{\mu\nu}$ is finite at $r=0$. The skew field $g_{[\mu\nu]}$ is like a
``medium"
which diffuses the spacetime metric and makes the non-Riemannian
geometry non-singular.
The ratio
\[
\epsilon=\frac{R_{R}}{R_{NR}},
\]
where $R_R$ and $R_{NR}$ denote the Riemannian and
non-Riemannian scalar
curvatures, behaves like a ``dielectric constant" in spacetime.

It is not meaningful to ask which geometry
is the ``correct" one, since the Riemannian and non-Riemannian
geometries
both play dual roles in the description of
spacetime. We shall adopt for convenience the definition of the line
element:
\begin{equation}
ds^2=g_{(\mu\nu)}dx^\mu dx^\nu.
\end{equation}

We will assume that $M \ll 1/\mu$. It can be shown that the only
solution which yields an asymptotically Minkowskian spacetime has
$w(r)=0$\cite{Clayton,Cornish1}.

In the case of the long-range approximation of NGT, corresponding to
$\mu\approx
0$ in the field equations, the exterior static spherically symmetric solution
has the
form (Wyman\cite{Wyman}):
\begin{mathletters}
\label{eq:exact_Wyman}
\begin{eqnarray}
\gamma_{\hbox{ext}}(r) &=& e^\nu, \\
\alpha_{\hbox{ext}}(r) &=& \frac{M^2
e^{-\nu}(1+s^2)}{(\cosh(a\nu)-\cos(b\nu))^2}
\left(\frac{d\nu}{dr}\right)^2, \\
\beta_{\hbox{ext}}(r) &=&r^2, \\
f_{\hbox{ext}}(r)&=& \frac{2M^2e^{-\nu}[\sinh(a\nu)\sin(b\nu)
+ s(1-\cosh(a\nu)\cos(b\nu))]}{(\cosh(a\nu)-\cos(b\nu))^2} ,
\end{eqnarray}
\end{mathletters}
where
\[
\begin{array}{ccc}
\displaystyle
a = \sqrt{\frac{\sqrt{1+s^2}+1}{2}} &
\hspace{.25in}\mbox{\rm and}\hspace{.25in} &
\displaystyle
b = \sqrt{\frac{\sqrt{1+s^2}-1}{2}}.
\end{array}
\]
$M$ is identified with the mass and $s$ is a dimensionless constant
which is different
for
different bodies and is
related to the strength of the coupling of matter to the skew field
$g_{[\mu\nu]}$.

The function $\nu(r)$ is determined by the relation:
\begin{equation}
\label{requation}
e^{\nu}[\cosh(a\nu)-\cos(b\nu)]^2\frac{r^2}{2M^2}
= \cosh(a\nu)\cos(b\nu) - 1 + s\sinh(a\nu)\sin(b\nu) .
\end{equation}

Two coordinate systems can be used to analyze the Wyman solution, one
of which
is the standard spherically symmetric coordinates: $x^1=r, x^2=\theta,
x^3=\phi, x^0=t$. Another useful coordinate system uses the coordinates:
$x^1=\nu, x^2=\theta, x^3=\phi, x^0=t$, where
\begin{equation}
\alpha_{\hbox{ext}}(\nu) = \frac{M^2 e^{-\nu}
(1+s^2)}{[\cosh(a\nu)-\cos(b\nu)]^2},
\end{equation}
\begin{equation}
\beta_{\hbox{ext}}(\nu) =
\frac{2M^2[\cosh(a\nu)\cos(b\nu)-1+s\sinh(a\nu)\sin(b\nu)]}
{e^{\nu}[\cosh(a\nu)-\cos(b\nu)]^2} ,
\end{equation}
and with $\gamma_{\hbox{ext}}(\nu)$ and $f_{\hbox{ext}}(\nu)$ given
as above.

We must choose a particular branch of a solution of Eq.(\ref{requation}).
This choice is made by picking the branch that will yield the
positive-mass Schwarzschild solution as a limit for $r\rightarrow\infty$.
This branch begins at $\nu=0$ and extends towards negative $\nu$.
In such a coordinate system, the asymptotic weak-field region is at
$\nu=0$ ($r=\infty$), while the ``origin'' occurs at $\nu_0$ defined by
$\beta(\nu_0)=0$.
The particular value of $\nu_0$ depends on the value of $s$, and
for $s=1$, we find numerically that $\nu_0\approx -5.1667$.

For $r\gg M$ and $|s|<1$, the metric functions in conventional spherical
coordinates are approximated by (for
$\mu\not=0$)\cite{CornMoff1,CornMoff2}:
\begin{mathletters}
\label{eq:app_larger_r}
\begin{eqnarray}
\gamma_{\hbox{ext}}(r) &\approx& 1 - \frac{2M}{r}, \\
\alpha_{\hbox{ext}}(r) &\approx& \left(1 - \frac{2M}{r}\right)^{-1}, \\
f_{\hbox{ext}}(r) &\approx& \frac{sM^2}{3}e^{-\mu r}(1+\mu r) .
\end{eqnarray}
\end{mathletters}
We see that for large $r$, the solutions for $\gamma_{\hbox{ext}}$ and
$\alpha_{\hbox{ext}}$ have asymptotically the same form as the
Schwarzschild vacuum solution in GR, and $f_{\hbox{ext}}(r)$ vanishes
exponentially fast as $r\rightarrow\infty$.

Near $r=0$ we can develop expansions where $r/M < 1$ and
$0 < \vert s\vert<1$. The leading terms are
\begin{eqnarray}
\gamma_{\hbox{ext}}(r)&=&\gamma_0+{\gamma_0(1+{\cal
O}(s^2))\over 2\vert
s\vert}\biggl({r\over M}\biggr)^2 + {\cal O}\biggl(\biggl({r\over
M}\biggr)^4\biggr),\\
\alpha_{\hbox{ext}}(r)&=&{4\gamma_0(1+{\cal O}(s^2))\over
s^2}\biggl({r\over
M}\biggr)^2
+{\cal O}\biggl(\biggl({r\over M}\biggr)^4\biggr),\\
f_{\hbox{ext}}(r)&=&M^2\biggl(4-{\vert s\vert\pi\over 2}+s\vert
s\vert+{\cal
O}(s^3)\biggr)
+{\vert s\vert+s^2\pi/8+{\cal O}(s^3)\over 4}r^2+{\cal O}(r^4),
\end{eqnarray}
where
\[
\gamma_0=\hbox{exp}\biggl(-{\pi+2s\over \vert s\vert}+{\cal
O}(s)\biggr)...\,.
\]

These solutions clearly illustrate
{\it the non-analytic nature of the limit} $s\rightarrow 0$ {\it in the strong
gravitational field regime} $0 < r \leq 2M\,$
\cite{CornMoff1,CornMoff2,Sokolov}.
Thus, the Wyman solution cannot be analytically continued to the
Schwarzschild solution of GR for arbitrarily small values of the parameter
$s$
in the region $0 < r\leq 2M$. This means that the expansion about a fixed
GR
background, in the approximation of linear $g_{[\mu\nu]}$, {\it cannot
be considered
valid in the strong gravitational regime}: $0 < r \leq 2M$. This fact will
play an important
role in our derivation of a solution to the collapse problem in NGT.

To see that the non-analytic behavior in $s$ of the Wyman solution for
$0<r\leq 2M$
is not a coordinate dependent result, we can use the coordinate invariant
norm:
\[
\sqrt{{}^{(t)}\xi^\mu{}^{(t)}\xi_\mu}=\gamma,
\]
where ${}^{(t)}\xi_\mu$ is the timelike Killing vector at spatial infinity:
\[
{}^{(t)}\xi_\mu=(\gamma,0,0,0).
\]
Since $\gamma$ never vanishes throughout the spacetime, there are no
event horizons
and the solution is not analytic to the Schwarzschild solution, for $0 < r
\leq 2M$
as $s\rightarrow 0$, in any coordinate frame. The redshift is finite for
$s\not=0$ and the
maximum redshift determined between $r=0$ and $r=\infty$ is given by
\[
z_{\hbox{max}}=\frac{1}{\sqrt{\gamma}}-1.
\]

The singularity caused by the vanishing of $\alpha(r)$ at
$r=0$ is a  coordinate singularity, which can be removed by
transforming to another coordinate frame of reference
\cite{CornMoff1,CornMoff2}.

The non-Riemannian geometry is non-singular in the range
$0\leq r < \infty$, since all the nonsymmetric curvature tensor invariants
are
finite in this range of $r$. For example, from the non-Riemmanian
curvature
tensor:
\[
{R^\lambda}_{\mu\nu\rho}(\Gamma)=\Gamma^\lambda_{\mu\nu,\rho}
-\Gamma^\lambda_{\mu\rho,\nu}
-\Gamma^\lambda_{\alpha\nu}\Gamma^\alpha_{\mu\rho}
+\Gamma^\lambda_{\alpha\rho}\Gamma^\alpha_{\mu\nu},
\]
we find that the Kretschmann invariant:
\[
K=R^{\lambda\mu\nu\rho}(\Gamma)R_{\lambda\mu\nu\rho}(\Gamma),
\]
where
\[
R_{\lambda\mu\nu\rho}(\Gamma)=
g_{\lambda\sigma}{R^\sigma}_{\mu\nu\rho}(\Gamma)
\]
is finite. On the other hand, the curvature invariants
formed from the
Riemann-Christoffel curvature tensor ${B^\lambda}_{\mu\nu\rho}$,
defined by
\[
{B^\lambda}_{\mu\nu\rho}=\left\{{\lambda\atop
\mu\nu}\right\}_{,\rho}-\left\{{\lambda\atop \mu\rho}\right\}_{,\nu}
-\left\{{\lambda\atop \alpha\nu}\right\}\left\{{\alpha\atop
\mu\rho}\right\}
+\left\{{\lambda\atop \alpha\rho}\right\}\left\{{\alpha\atop
\mu\nu}\right\}
\]
are singular like $\sim M^4/r^8$ near $r=0$\cite{Moffat3,Cornish2}.
The Riemann-Christoffel
curvature tensors built out of the two metrics ${}^2m$ and ${}^3m$ will
also have this singular
type of behavior near $r=0$.

In the $\nu$ coordinate representation, $\gamma_{\hbox{ext}}(\nu)$
vanishes at the point $\nu=-\infty$, corresponding to
$r^*=(r^4+f^2)^{1/4}=0$,
which is a point outside the physical causal spacetime.
Moreover, there is a singularity in the non-Riemannian curvature
invariants at the
unphysical point $\nu=-\infty$\cite{Cornish2}.

Let us consider the exterior static spherically symmetric line element
given by the metric
${}^1m_{(\mu\nu)}=g_{(\mu\nu)}$:
\begin{equation}
ds^2=\gamma_{\hbox{ext}}(\nu)dt^2-\alpha_{\hbox{ext}}(\nu)d\nu^2-
\beta_{\hbox{ext}}(\nu)(d\theta^2+\sin^2\theta d\phi^2).
\end{equation}
We see that $\gamma_{\hbox{ext}}(\nu)$ does not vanish in the range:
$-\infty < \nu \leq 0$.
Thus, all the points of spacetime are described
by timelike Killing vector fields in the region
$-\infty < \nu \leq 0$; they can be causally connected and all particles can
be at rest.
This corresponds to the fact that there are no trapped surfaces (event
horizons) in this
range of $\nu$. There are no event horizons in the $r$-coordinate frame
in the region: $0 < r \leq  \infty$.

Test particles which follow radial trajectories according to the path
equation\cite{Moffat3}:
\[
\frac{du^\mu}{ds}+\Gamma^\mu_{\alpha\beta}u^\alpha u^\beta=0
\]
or test particles that follow radial trajectories according to the
geodesic equation, Eq.(\ref{geodesic}),
are not stopped at $r=0$ but can continue through into the unphysical
vacuum manifold for $r < 0$\cite{LegareMoff2}. However, we shall see
that
in the collapse of physical bodies, it is unlikely that matter can collapse
through the
point $r=0$ and we conjecture that this is never possible.

\section{Matching of Interior and Exterior Solutions}

We must consider now the matching of the metric outside the star with
the one in the
interior of the star. Birkhoff's theorem does not hold for the spherically
symmetric
vacuum solution of NGT, which makes the matching of solutions a more
difficult task
to solve than in GR. It can be proved for $\mu\not=0$ that no monopole
radiation can escape to
asymptotic infinity\cite{Clayton}.

The exterior metric outside the star can be expressed in terms of the
coordinates,
${\bar r}, {\bar \theta}, {\bar \phi}, {\bar t}$ in the form:
\begin{equation}
ds^2=C({\bar r},{\bar t})d{\bar t}^2-D({\bar r},{\bar t})d{\bar r}^2
-2E({\bar r},{\bar t})d{\bar r}d{\bar t}-{\bar r}^2
(d{\bar \theta}^2+\sin^2{\bar \theta}d{\bar \phi}^2).
\end{equation}
We remove $E$ by defining the new time variable:
\begin{equation}
d{\bar t}'=\epsilon({\bar r},{\bar t})[C({\bar r},{\bar t})d{\bar t}-
E({\bar r},{\bar t})d{\bar r}].
\end{equation}
The line element now takes the form\cite{Weinberg}:
\begin{equation}
ds^2=\gamma_{\hbox{ext}}({\bar r},{\bar t})d{\bar t}^2
-\alpha_{\hbox{ext}}({\bar r},{\bar t})d{\bar r}^2
-{\bar r}^2(d{\bar \theta}^2+\sin^2{\bar \theta}d{\bar \phi}^2),
\end{equation}
where
\begin{eqnarray}
\gamma_{\hbox{ext}}({\bar r},{\bar t})&=&\epsilon^{-2}({\bar r},{\bar
t})
C^{-1}({\bar r},{\bar t}),\\
\alpha_{\hbox{ext}}({\bar r},{\bar t})&=&D({\bar r},{\bar
t})+C^{-1}({\bar r},{\bar t})
E^2({\bar r},{\bar t}).
\end{eqnarray}
Let us first consider the approximation scheme in which
Eq.(\ref{approx1})
is valid. This approximation is expected to hold for the initial
phase of the collapse when $f$ is small. In order to match the solutions at
the surface
of the star, we must convert the interior solution with the metric
(\ref{FRWmetric}) into
``standard'' coordinates.  We shall assume that in the initial phase of the
collapse, the interior metric is approximately determined by
the solution Eq.(\ref{FRW}). We choose~\cite{Weinberg}
\begin{equation}
{\bar r}=rR(t),\quad {\bar \theta}=\theta,\quad {\bar \phi}=\phi,
\end{equation}
and use an integrating factor which yields
\begin{equation}
{\bar t}=\biggl(\frac{1-kr_0^2}{k}\biggr)\int^1_{Q(r,t)}
\frac{dR}{(1-\frac{kr_0^2}{R^2})}
\biggl(\frac{R}{1-R}\biggr)^{1/2},
\end{equation}
where
\[
Q(r,t)=1-\biggl(\frac{1-kr^2}{1-kr_0^2}\biggr)^{1/2}(1-R(t)).
\]
Here the constant $r_0$ is set equal to the radius of the star in comoving
coordinates.
We now get
\begin{eqnarray}
\gamma({\bar r},{\bar
t})&=&\frac{R}{Q}\biggl(\frac{1-kr^2}{1-kr_0^2}\biggr)^{1/2},\\
\alpha({\bar r},{\bar t})&=&\biggl(1-\frac{kr^2}{R}\biggr)^{-1},
\end{eqnarray}
where $Q$ and $R$ are functions of ${\bar r}$ and ${\bar t}$. At the
radius of the
star, we have
\begin{eqnarray}
{\bar r}&=&{\bar r}_0=r_0R(t)\\
{\bar t}&=&\biggl(\frac{1-kr_0^2}{k}\biggr)^{1/2}
\int^1_{R(t)}\frac{dR}{\biggl(1-\frac{kr_0^2}{R}\biggr)}\biggl(\frac{R}{1-R}
\biggr)^{1/2},\\
\gamma({\bar r_0},{\bar t})&=&1-\frac{kr_0^2}{R(t)},\\
\alpha({\bar r_0},{\bar t})&=&\biggl(1-\frac{kr_0^2}{R(t)}\biggr)^{-1}.
\end{eqnarray}

The exterior and interior solutions match at ${\bar r}=r_0R(t)$ if we have
\begin{mathletters}
\begin{eqnarray}
\label{match1}
\gamma({\bar r}_0,{\bar t})&=&\gamma_{\hbox{ext}}({\bar r}_0,{\bar
t}),\\
\label{match2}
\alpha({\bar r}_0,{\bar t})&=&\alpha_{\hbox{ext}}({\bar r}_0,{\bar t}),
\\
\label{match3}
f({\bar r}_0,{\bar t})&=&f_{\hbox{ext}}({\bar r}_0,{\bar t}).
\end{eqnarray}
\end{mathletters}
We shall now expand the exterior time dependent solution as
\begin{mathletters}
\begin{eqnarray}
\gamma_{\hbox{ext}}({\bar r},{\bar
t})&=&\gamma_{\hbox{ext}}({\bar r})
+\delta\gamma_{\hbox{ext}}({\bar r},{\bar t}),\\
\alpha_{\hbox{ext}}({\bar r},{\bar t})&=&\alpha_{\hbox{ext}}({\bar
r})
+\delta\alpha_{\hbox{ext}}({\bar r},{\bar t}),\\
f_{\hbox{ext}}({\bar r},{\bar t})&=&f_{\hbox{ext}}({\bar r})
+\delta f_{\hbox{ext}}({\bar r},{\bar t}).
\end{eqnarray}
\end{mathletters}
We assume that
$\delta\gamma_{\hbox{ext}},\delta\alpha_{\hbox{ext}}$ and
$\delta f_{\hbox{ext}}$ are small quantities that can be neglected during
the collapse (quasi-static approximation).
The $\alpha_{\hbox{ext}}({\bar r}), \gamma_{\hbox{ext}}({\bar r})$
and
$f_{\hbox{ext}}({\bar r})$ are determined by the unique static solution
of the NGT
vacuum field equations.

For large values of the star's surface radius, $r_0 >>2M$, we have
\begin{eqnarray}
\alpha_{\hbox{ext}}(r_0)&=& \biggl(1-\frac{2M}{r_0}\biggr)^{-1},\\
\gamma_{\hbox{ext}}(r_0)&=&1-\frac{2M}{r_0},\\
\label{fasymptotic}
f_{\hbox{ext}}(r_0)&=&\frac{sM^2}{3}\exp(-r_0/a)(1+r_0/a).
\end{eqnarray}
If, as we have been assuming, $r_0 \ll a$, then we get from
(\ref{fasymptotic}):
\begin{equation}
f_{\hbox{ext}}(r_0)=\frac{sM^2}{3}.
\end{equation}
Let us assume that $s$ is related to the strength of the skew field coupling
to matter by
the expression:
\begin{equation}
s=\frac{g}{N_B^\beta},
\end{equation}
where $g$ is a coupling constant, $N_B$ is the baryon number of the star
and $\beta$ is some
dimensionless constant.  We expect that for a non-zero coupling of the
skew field
to matter there should exist a non-vanishing static exterior solution for
$f$.

The interior and exterior solutions fit for large $r_0$ when
${\bar r}=r_0R(t)$ if:
\begin{equation}
kr_0^2=\frac{2M}{r_0}.
\end{equation}
This yields for large $r_0$ the expression for the total mass of the star:
\begin{equation}
M=\frac{4\pi}{3}\rho(0)r_0^3.
\end{equation}

For $r_0R(t)\rightarrow 2M$, we must switch to the approximation
scheme
based on the condition (\ref{approx2}), since the quasi-static
Schwarzschild
solution fails to be a solution of the NGT field equations for $r_0R \leq
2M$\cite{Sokolov}, provided we make the reasonable assumption that
the limit to the static solution from small time dependent perturbations is
smooth. We assume as before that the
solution can be expressed as a separation of variables, as in
Eq.(\ref{sepvariables}).
Then the line element takes the normal Gaussian form:
\begin{equation}
ds^2=dt^2-R^2(t)[q(r)dr^2+r^2(d\theta^2+\sin^2\theta d\phi^2)].
\end{equation}
To express this line element in standard form, we must use the integration
factor:
\begin{equation}
{\bar t}'=\int dR{\cal F}(R,r_0),
\end{equation}
where ${\cal F}$ is a function that is chosen to remove the cross-term
$drdt$. The matching of the exterior and interior solutions is achieved by
use
of the conditions (\ref{match1})-(\ref{match3}).

Because there are no trapped surfaces (event horizons) in the exterior
Wyman
solution in the range $0 < {\bar r} \leq\infty$, we have
\begin{equation}
\label{noblackhole}
\gamma_{\hbox{ext}}(2M)\not=0,\quad
\alpha_{\hbox{ext}}(2M) < \infty,\quad
f_{\hbox{ext}}(2M)\not=0.
\end{equation}
{}From this we can deduce that the matching of the interior and the exterior
solutions will {\it not produce a black hole event horizon in the final stage
of collapse}.

A light signal emitted in a radial direction at ${\bar t}$ will have $d{\bar
r}/d{\bar t}$
determined by $ds^2=0$ where
\[
ds^2=\gamma({\bar r},{\bar t})dt^2-\alpha({\bar r},{\bar t})d{\bar r}^2
-{\bar r}^2(d\theta^2+\sin^2\theta d\phi^2).
\]
The light signal will be detected at a distant point ${\bar r}$ at a time:
\[
{\bar t}'={\bar t}+\int_{r_0R(t)}^{{\bar r}'} d{\bar r}
\biggl[\frac{\alpha_{\hbox{ext}}({\bar r})}{\gamma_{\hbox{ext}}({\bar
r})}
\biggr]^{1/2}.
\]
Because of the bounds in (\ref{noblackhole}), we find that
${\bar t}' <\infty$ and the collapse to the Schwarzschild radius can occur
in a {\it finite} time.

In GR, we have
\[
{\bar t}'\sim
-2kr_0^3\ln\biggl[1-\frac{kr_0^2}{R(t)}\biggr]+\hbox{const},
\]
so that as $r_0R\rightarrow 2M=kr_0^3$, it takes an infinite amount of
time
for the star to collapse to the Schwarzschild radius.

For large $r_0$, we have
\[
\frac{kr^2}{R(t)} << 1
\]
and the redshift is given by
\[
z\sim r_0\sqrt{k}\biggl(\frac{1-R(t)}{R(t)}\biggr)^{1/2}
\sim r_0\sqrt{k}\biggl(1-R({\bar t}'-{\bar r}')(R({\bar t}'-{\bar
r}')\biggr)^{1/2},
\]
which is asymptotically the same as the result in GR.

As $r_0\rightarrow 2M$, the redshift is determined by the formula:
\begin{equation}
z\equiv \frac{\lambda'-\lambda_0}{\lambda_0}=\frac{d{\bar t}'}{dt}
-r_0\dot {R}(t)
\biggl[\frac{\alpha_{\hbox{ext}}(r_0R)}{\gamma_{\hbox{ext}}(r_0R)}
\biggr]^{1/2}-1.
\end{equation}
The red shift $z$ seen by an observer is zero when the collapse is
observed
to begin, increases slowly during the collapse but satisfies the bound
$z < \infty$, because Eq.(\ref{noblackhole}) holds for $0 < r_0R(t) <
\infty$.

In GR, the redshift becomes for $r_0R(t)\rightarrow 2M$:
\begin{equation}
z\sim 2\biggl(1-\frac{kr_0^2}{R(t)}\biggr)^{-1}\sim
\exp\biggl(\frac{{\bar t}'}{2kr_0^3}\biggr),
\end{equation}
and the red shift becomes infinitely large as the collapsing star approaches
the black hole event horizon.

Since a black hole event horizon does not form in the collapse of a
star, in NGT, if we formulate the collapse problem with our matching
conditions, the star is not cut off from the rest of the universe,
and it can continue to emit all forms of radiation.
The collapsed star is expected to form a dense static object
that could be a strong source of gravitational radiation and other forms of
radiation.

\section{Concluding Remarks}

We have been able to find an approximate treatment, in NGT, of the
stellar equilibrium problem. We found that it is possible to realize
a static equilibrium state for a massive compact object, due to the
strong repulsive forces at the center of the object produced by the
skew fields. We also formulated an approximate treatment of the
gravitational
collapse of a star considered as a spherically symmetric, pressureless
dust sphere. By
matching the exterior vacuum solution to the interior solution,
we have shown that black holes are not expected to form during the
collapse,
since the exterior solution for small time dependent
perturbations can approximate the static Wyman solution, which does
not possess any trapped surfaces. The collapse can be
stopped before the center of the star is reached.
To completely solve the problem of the collapse of a star, in NGT,
we must find a solution of the full NGT
field equations, preferably for $\mu\not=0$ and including a suitable
$K^{[\mu\nu]}$
source contribution. Such a solution is expected to be found only by using
numerical
methods to solve the field equations.

It is important to stress that the
collapse of a star, in NGT, must be stopped before $R(t)=0$ is reached,
since the
spacetime becomes unphysical for $R(t) < 0$. Because of the absence of
black hole event horizons, such unphysical behavior would be ``naked"
and it
would destroy the physical Cauchy data and make NGT a non-viable
classical theory of gravitation. Of course, it is possible that we would
have to discover a quantum theory of gravity to fully comprehend the
collapse to small distances. But since such a theory is not presently
available, it would be more satisfactory to be able to settle the issue
of small distance collapse within the classical regime of NGT.

It has been conjectured by Burko and Ori~\cite{Burko} that black holes
should be anticipated in the gravitational collapse of stars in NGT.
Their analysis relied entirely on the use of an expansion of the NGT time
dependent
field equations about a GR background to first order in small
$g_{[\mu\nu]}$.
Specifically, this background was chosen to be the static spherically
symmetric
Schwarzschild solution. Of course, with this assumption a black hole is
expected to
form, because of the existence of a no-hair theorem for a skew symmetric
potential
coupled to a Schwarzschild metric\cite{CornMoff3}, and the existence of
trapped surfaces in the Schwarzschild solution.

Since Burko and Ori demanded that $f$ be small throughout the collapse
of
a star, they used a quasi-static exterior metric that can be closely
approximated by the static
Schwarzschild solution for small enough $f$. This metric {\it cannot} be
a solution
of the NGT vacuum field equations for $0 < r \leq 2M$\cite{Sokolov}.
The exact static spherically symmetric vacuum solution, in NGT, is the
Wyman
solution which does not have event horizons for $0 < r \leq\infty$. It is
{\it not}
analytic to the Schwarzschild solution for arbitrarily small values of the
parameter $s$
in the range $0 < r \leq 2M$.
{}From this we conclude that a quasi-static exterior metric is expected to be
the Wyman
metric plus a small time dependent part, which can be valid for suitably
chosen
initial value data with small $f$, before the onset of collapse. When the
interior
solution is matched to this exterior solution, black holes are not expected
to
form during collapse.
This invalidates the claim made by Burko and Ori that
black holes can be anticipated in the collapse of a star in NGT.

It is incorrect to claim any definitive results for non-linear gravity
theories, such as
GR and NGT, on the basis of the linear approximation. One would not
use
the linear approximation in GR to solve the collapse problem, because if
we expand
the metric about Minkowski space, then for $r=2M$, the perturbative
expansion fails to
be valid, since the metric perturbation is of order 1.  Similarly, the
Burko-Ori quasi-static
expansion of $g_{\mu\nu}$ about the
Schwarzschild background fails to hold at $r=2M$, because
$g_{[\mu\nu]}$ becomes
larger than unity and the linear approximation breaks down. Given that a
generic source
exists that generates a non-zero $g_{[\mu\nu]}$, a static solution for $f$
is produced. Then,
we know from the exact Wyman solution that $f$ becomes large near
$r=2M$, and the
linear equation for $f$ fails to describe correctly the physical collapse of a
star.
Assuming that $f$ is just a radiating wave without a static part during the
collapse is
not a physically realistic treatment of the problem in NGT. Burko and Ori
incorrectly assumed that by adding matter and a generic coupling of
matter
to the skew fields for gravitational collapse would not change their
conclusions.

The NGT violates the strong equivalence principle and, therefore, it is
expected that a freely falling
observer (the weak equivalence principle is not violated in the new
massive version of NGT)
could perform experiments to
detect the formation of an event horizon. This is not possible in GR. This
effect would
only show itself in higher orders of $g_{[\mu\nu]}$, and would have
important
consequences for the physics of collapse in NGT.

Actual stars would collapse more slowly than
in the model which we have studied because of the effect of the
pressure of radiation, of matter and of rotation.

Since the final collapsed object is expected to be a massive compact
star without an event horizon, radiation of all forms can be emitted by the
surface of the star. Of course, if the red shift emitted from the surface of
the star
is too large (but never infinite), then in practice only small amounts of
thermal and
gravitational radiation will escape. There would be no Hawking radiation
emitted, for such
radiation is associated specifically with a rigorous black hole event
horizon. Therefore, the
problem of information loss associated with the quantum mechanical
aspects of a black hole
would be eliminated.

\acknowledgements

This work was supported by the Natural Sciences and Engineering
Research Council of Canada. I thank M. A. Clayton, N. J. Cornish,
L. Demopoulos, J. L\'egar\'e, M. Reisenberger and P. Savaria for helpful
discussions.

\appendix

\section{The Time Dependent $\Gamma$-connections}

The NGT compatibility equation is given by
\begin{equation}
g_{\lambda\nu,\eta}-g_{\rho\nu}\Gamma^\rho_{\lambda\eta}
-g_{\lambda\rho}\Gamma^\rho_{\eta\nu}=\frac{1}{6}g^{(\mu\rho)}
(g_{\rho\nu}g_{\lambda\eta}-g_{\eta\nu}g_{\lambda\rho}-
g_{\lambda\nu}g_{[\rho\eta]})W_\mu,
\end{equation}
where $W_\mu$ is determined from (\ref{Wequation}). For the
spherically
symmetric system, when $w(r,t)=g_{[01]}(r,t)=0$,
it follows that $W_\mu=0$ and the compatibility equation reads:
\begin{equation}
g_{\lambda\nu,\eta}-g_{\rho\nu}\Gamma^\rho_{\lambda\eta}
-g_{\lambda\rho}\Gamma^\rho_{\eta\nu}=0.
\end{equation}

The non-vanishing components of the $\Gamma$-connections are:
\begin{eqnarray}
\Gamma^1_{11}&=&\frac{\alpha'}{2\alpha},\\
\Gamma^1_{(10)}&=&\frac{\dot{\alpha}}{2\alpha},\\
\Gamma^1_{22}&=&\Gamma^1_{33}\hbox{cosec}^2\theta
=\frac{1}{2\alpha}\biggl(fB-\frac{1}{2}\beta A'\biggr),\\
\Gamma^1_{00}&=&\frac{\gamma'}{2\alpha},\\
\Gamma^2_{(12)}&=&\Gamma^3_{(13)}=\frac{1}{4}A',\\
\Gamma^2_{(20)}&=&\Gamma^3_{(30)}=\frac{1}{4}\dot{A},\\
\Gamma^2_{33}&=&-\sin\theta\cos\theta,\\
\Gamma^3_{(23)}&=&\cot\theta,\\
\Gamma^0_{(11)}&=&\frac{\dot{\alpha}}{2\gamma},\\
\Gamma^0_{(10)}&=&\frac{\gamma'}{2\gamma},\\
\Gamma^0_{22}&=&\Gamma^0_{33}\hbox{cosec}^2\theta
=- \frac{1}{2\gamma}\biggl(fD-\frac{1}{2}\beta\dot{A}\biggr),\\
\Gamma^0_{00}&=&\frac{\dot{\gamma}}{2\gamma},\\
\Gamma^1_{[23]}&=&\frac{\sin\theta}{2\alpha}\biggl(\frac{1}{2}fA'+\beta
B\biggl),\\
\Gamma^2_{[13]}&=&-\Gamma^3_{[12]}\sin^2\theta=\frac{1}{2}B\sin\theta,\\
\Gamma^2_{[30]}&=&- \Gamma^3_{[20]}\sin^2\theta=-
\frac{1}{2}D\sin\theta,\\
\Gamma^0_{[23]}&=&-
\frac{\sin\theta}{2\gamma}\biggl(\frac{1}{2}f\dot{A}
+\beta D\biggl),
\end{eqnarray}
where $A, B$ and $D$ are given by Eqs.(\ref{Aequation}),
(\ref{Bequation}) and
(\ref{Dequation}).

\end{document}